# Genuine Dirac half-metal: A 2D $d^0$-type ferromagnet Mg$_4$N$_4$

Jialin Gong[a,1], Guangqian Ding[b,1], Chengwu Xie[a], Jianhua Wang[a], Ying Liu[c,*], Xiaotian Wang[a,d,*], Gang Zhang[e,*]

[a]School of Physical Science and Technology, Southwest University, Chongqing 400715, China;

[b]School of Science, Chongqing University of Posts and Telecommunications, Chongqing 400065, China;

[c] School of Materials Science and Engineering, Hebei University of Technology, Tianjin 300130, China

[d]Institute for Superconducting and Electronic Materials (ISEM), University of Wollongong, Wollongong 2500, Australia

[e]Institute of High Performance Computing, Agency for Science, Technology and Research (A*STAR), 138632 Singapore

[*]Corresponding authors.

E-mail addresses: ying_liu@hebut.edu.cn; xiaotianwang@swu.edu.cn, zhangg@ihpc.a-star.edu.sg

**Abstract**

When the spin-orbit coupling (SOC) is absent, almost all the proposed half-metals with the twofold degenerate nodal points at the K (or K') in two-dimensional (2D) materials are misclassified as "Dirac half-metals" owing to the way graphene was utilized in the earliest studies. Actually, each band crossing point at K or K' is described by a 2D Weyl Hamiltonian with definite chirality; hence, it must be a Weyl point. To the best of our knowledge, there have been no reports of a genuine (*i.e.*, fourfold degenerate) Dirac point half-metal in 2D yet. In this Letter, we proposed for the first time that the 2D $d^0$-type ferromagnet Mg$_4$N$_4$ is a genuine Dirac half-metal with a fourfold degenerate Dirac point at the S high-symmetry point, intrinsic magnetism, high Curie temperature, 100% spin-polarization, robustness to the SOC and uniaxial and biaxial strains, and 100% spin-polarized edge states. The work can be seen as a starting point for future predictions of intrinsically magnetic materials with genuine Dirac points, which will aid the frontier of topo-spintronics researchers.

The appearance of intrinsic magnetism in two-dimensional (2D) crystals [1-5] is a significant recent accomplishment in materials science, as it opens the door to new cutting-edge disciplines in the 2D family and could lead to revolutionary data storage and information systems with further downsizing. Nevertheless, the majority of 2D materials [6], including graphene, lack intrinsic magnetism, limiting their practical utility as spintronic devices. In recent years, a variety of advanced techniques [7], including doping, defects, functionalization, *etc.*, have been used to create magnetic in graphene-based 2D materials. Unfortunately, these materials' magnetism is insufficiently robust, which hinders their practical implementation as spintronic devices [8]. The intrinsic magnetic properties of 2D materials have been the subject of numerous investigations. A number of ferromagnetic half-metals have been theoretically predicted and actually synthesized [9-11], including 2D $CrI_3$ monolayers, $Cr_2Ge_2Te_6$ bilayers, and $Fe_3GeTe_2$ monolayers.

Furthermore, investigating the link between magnetism and band topology in the 2D family has become a popular study topic [12-30]. As indicated in Table S1, we reviewed a series of published literature and discovered that many researchers have proposed 2D half-metals (or 2D spin-gapless semiconductors) with linearly dispersing crossing points. We would like to point all the linearly dispersing crossing points with twofold degeneracy in the mentioned literature (see Table S1) are misclassified as "Dirac points".

This is likely a result of the use of graphene in early research. Each band crossing point at the K or K' in graphene is characterized by a 2D Weyl Hamiltonian with a definite chirality, and hence should be a Weyl point. For concreteness, the effective model describing the band crossings at K or K' is given as $\mathcal{H}_{K/K'} = v(\tau_z k_x \sigma_x + k_y \sigma_y)$, with $v$ as the Fermi velocity, and $\sigma's$ stand for the Pauli matrix. Here, $\tau = \pm 1$, refers to K and K' valley, respectively. $\mathbf{k}$ is measured from the band crossings at K/K'. Based on the effective model, the eigenenergies for the crossing bands are $E_{c/v} = \pm v\sqrt{k_x^2 + k_y^2}$, implying a double degeneracy at K/K' point. Hence, the degeneracy for the band crossings at the corners of graphene Brillouin zone is double, rather than a fourfold degeneracy. Tranditionally, we call the band crossings in graphene as "Dirac" point is because we combine the quasiparticles at these two valleys into a single model. However, the degeneracy does not occurs at the same point. Therefore, it is a 2D Weyl point. That is to say, it is conceptually more coherent to refer to the aforementioned materials (listed in Table S1 in the SM) as Weyl half-metals with Weyl points (twofold degenerate points that satisfy the Weyl model) as opposed to Dirac half-metals with Dirac points (fourfold degenerate points that satisfy the Dirac model). Moreover, in most proposed 2D Weyl half-metals/spin-gapless semiconductors, small gaps between two Weyl points (usually at the K and K') occur when the SOC is considered. The opening of the gap for a 2D Weyl point may not be a bad thing; instead, it may generate intriguing physics, such as the quantum anomalous Hall effect (QAHE) in 2D systems [23,31].

Unfortunately, besides 2D Weyl half-metals/spin-gapless semicondu,ctors with twofold degenerate Weyl points around/at the Fermi level, no genuine 2D Dirac half-metals/spin-gapless semiconductors with fourfold degenerate Dirac points around/at the Fermi level can be found in the literature. *Consequently, the question becomes whether a genuine 2D Dirac half-metal with a fourfold degenerate Dirac point is possible.* The answer is yes. Although the 2D Dirac points have not been observed in 2D magnetic systems, they have been postulated in nonmagnetic semiconductors and phonon curves of 2D materials [32-34].

Possible explanations for why 2D Dirac points have not been studied in 2D magnetic systems include: (1) In spintronic, linearly dispersing band crossing points in 2D are almost always referred to as "Dirac points", regardless of the degeneracy of the bands surrounding the band crossing points [35]. This indicates that many researchers fail to differentiate between the Weyl and Dirac points, particularly in two-dimensional magnetic system materials. Consequently, the prediction of genuine Dirac half-metals in 2D has not received sufficient consideration. (2) The search for intrinsic 2D magnetic materials is ongoing, let alone predicting a genuine 2D Dirac half-metal with intrinsic magnetism and 100% spin-polarization. (3) Two issues also plague 2D topological ferromagnets. The first [22,25] is that most topological signatures in 2D topological ferromagnets will be gapped under spin-orbit coupling. The other is the current 2D topological ferromagnets' relatively low Curie temperature. As a result, finding 2D Dirac half-metals with a high Curie temperature and robust spin-orbit coupling may be difficult.

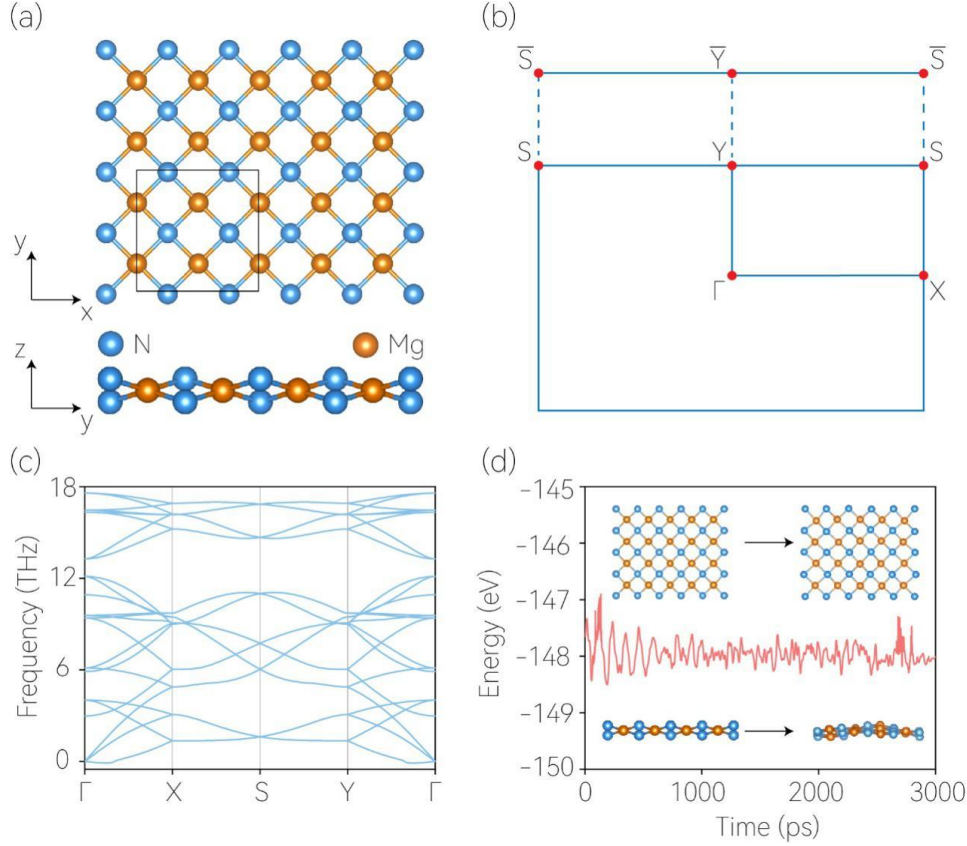

Figure 1. (a) Top and side views of $Mg_4N_4$ monolayer. The dashed box marks the unit cell. (b) 2D BZ and its projection to the [010] edge. (c) The calculated phonon dispersion for 2D $Mg_4N_4$ unit cell. To obtain the force constants, we adopt a $2 \times 2 \times 1$ supercell of $Mg_4N_4$. (d) Ab-initio molecular dynamics of $Mg_4N_4$ at 300K.

In this Letter, we proposed the first Dirac half-metal in 2D ferromagnetic material $Mg_4N_4$, with a fully spin-polarized fourfold degenerate Dirac point at the S high-symmetry point, a high Curie temperature of about 431 K, visible fully spin-polarized edge state, and the robustness of the spin-orbit coupling and the uniaxial and biaxial strains. Moreover, the magnetic moments in existing 2D topological ferromagnets are primarily contributed by unoccupied $d/f$ shells from transition-metal or rare-earth elements, which is known as $d/f$-type ferromagnetism. However, the magnetism in the $Mg_4N_4$ example arises from the $p$ orbitals. That is, $Mg_4N_4$ is a 2D monolayer without unoccupied d/ f shells, namely, a $d^0$ ferromagnet [14,36-38]. There have been fewer studies on $d^0$ ferromagnets than on 2D $d/f$-type ferromagnets, and the $d^0$ ferromagnets have unique properties, such as much weaker localization of $p$ electrons and smaller spin-orbit coupling strength, which are significant advantages for high-speed and long-distance transport [36,37].

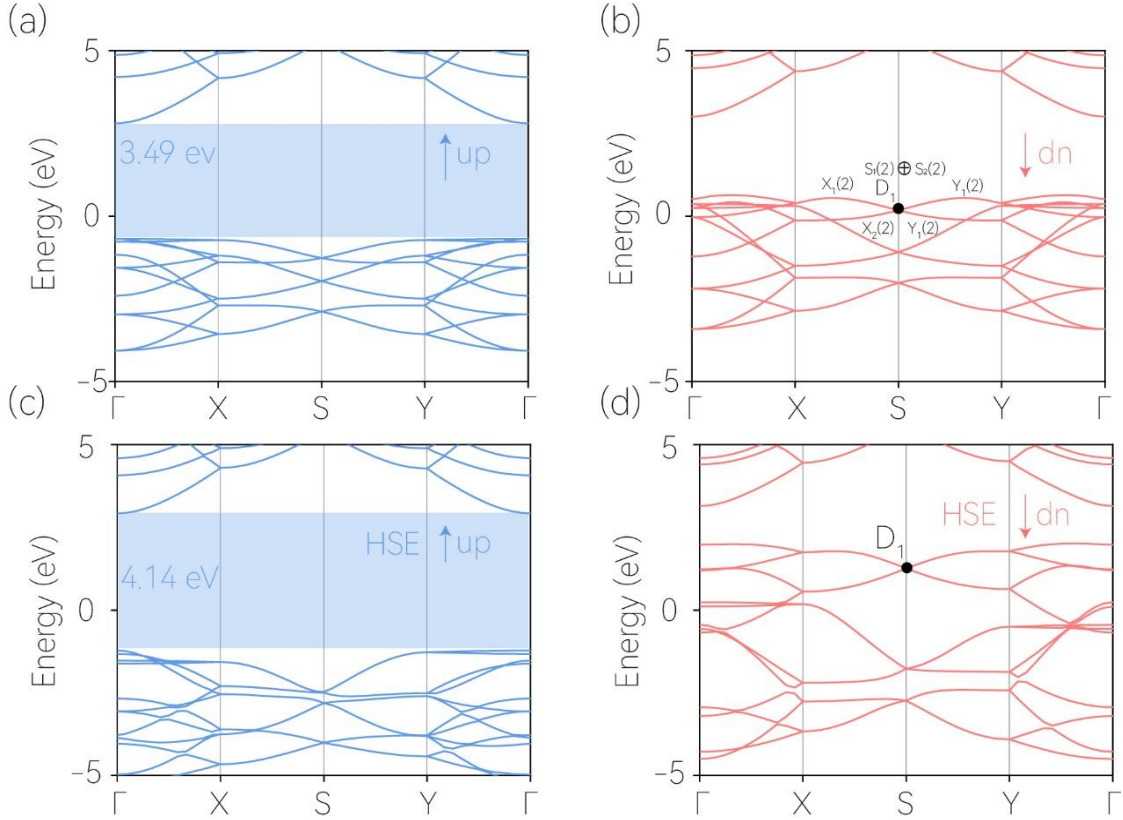

Figure 2. (a) and (b) spin-up and spin-down band structures for 2D $Mg_4N_4$ calculated by the GGA method. (c) and (d) spin-up and spin-down band structures for 2D $Mg_4N_4$ calculated by the HSE 06 method. The $D_1$ shows the fully spin-polarized Dirac point in the spin-down channel.

In the main text, we select the 2D $Mg_4N_4$ with layer group No. 45 (space group No. 57) as a typical example, the lattice structure of the monolayer $Mg_4N_4$ is displayed in Figure 1a. More details about the information on atomic positions for $Mg_4N_4$ can be found in Table S2. Before investigating its electronic band structures, the dynamical and thermal stability of the lattice should be confirmed. We used the density functional perturbation theory [39] to obtain the force constants, as implemented in the VASP. Then, we used the PHONOPY package [40] to calculate the phonon dispersion spectrum. Figure 1c shows no imaginary frequency in the phonon spectrum, showing that 2D $Mg_4N_4$ is dynamically stable. We perform the *ab initio* molecular dynamics (AIMD) [41] calculated in a 2 × 2 × 1 supercell to estimate the thermal stability for 2D $Mg_4N_4$. As shown in Figure 1d, after 3000 steps at 300 K, we find that thermal-induced fluctuations without bond breakage or geometric reconfigurations exist in the final states, demonstrating its thermally stable at room temperature.

Notice that 2D $Mg_4N_4$ belongs to a $d^0$ ferromagnet without unoccupied d/ f shells. In our calculations, we have considered four possible magnetic configurations, including one nonmagnetic, one ferromagnetic, and two antiferromagnetic states, for the unit cell (see Figure S1 in the SM). To identify the ground magnetic state of the $Mg_4N_4$ monolayer, we compare their energies and found that the ferromagnetic state has the lowest energy and should be the ground magnetic state for the 2D $Mg_4N_4$. In addition, six magnetic configurations in the 2×2×1 supercell, including one nonmagnetic, one ferromagnetic, and four antiferromagnetic states, were investigated further to confirm the magnetic ground state of 2D $Mg_4N_4$ (see Figure S2 in the SM). Clearly, the ferromagnetic state for $Mg_4N_4$ remains the lowest magnetic state. The determined total and atomic magnetic moments for 2D $Mg_4N_4$ are shown in Table S3. Table S3 reveals that practically all magnetic moments are concentrated around the N atoms, indicating that the magnetic character of $Mg_4N_4$ is primarily due to the N atoms. The projected density of states in Figure S3b shows that the low-energy bands are mainly from the N-$p$ orbital.

Also, we plot the electronic band structure of 2D $Mg_4N_4$ without considering the spin-orbit coupling in Figure 2. The band structures obtained from the GGA [42] in the spin-up and spin-down channels are shown in Figure 2a and Figure 2b, respectively. Moreover, the more accurate but computationally expensive Heyd–Scuseria–Ernzerhof (HSE 06) hybrid functional [43] is further employed to double-check the spin-polarized band structures of 2D $Mg_4N_4$, as exhibited in Figure 2c and Figure 2d, respectively. It can be noticed that spin-up and spin-down states are separated, with spin-down states contributing primarily to the states at the Fermi level. The spin-polarization ratio can be calculated via the formula as follows [44,45]:

$$P\ (\%) = \frac{N\uparrow(E_F) - N\downarrow(E_F)}{N\uparrow(E_F) + N\downarrow(E_F)}, \tag{1}$$

where $N\uparrow(E_F)$ and $N\downarrow(E_F)$ are the spin-up and spin-down electrons at the Fermi level ($E_F$), respectively. One finds that $N\uparrow(E_F) = 0$, and thus 2D $Mg_4N_4$ should be a half-metal with 100% spin polarization around $E_F$.

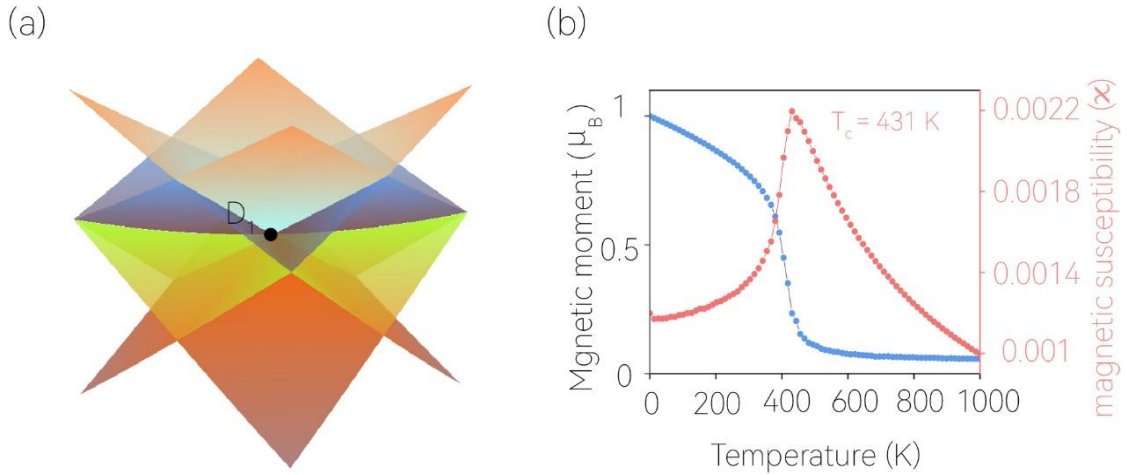

Figure 3. (a) The 3D plot of the four bands around the Dirac point $D_1$ at the S high-symmetry point. (b) Dependence of magnetic moment and magnetic susceptibility on the temperature by the Heisenberg model via Monto Carlo (MC) simulation.

Near the Fermi level, one can find that the two bands (from the spin-down channel) along the X-S and Y-S high-symmetry paths are twofold degeneracy. These two doubly degenerate bands along X-S and Y-S high-symmetry paths belong to Weyl nodal lines, which can be understood from the symmetry analysis as follows. LG 45 is generated by two screw rotation symmetry, namely, $S_{2x}: (x, y, z) \to (x + 1/2, -y + 1/2, -z)$, and $S_{2y}: (x, y, z) \to \left(-x + \frac{1}{2}, y, -z + \frac{1}{2}\right)$, and inversion symmetry $\mathcal{P}$. In the absence of SOC, the two spin channels are decoupled, each spin states could be regarded as spinless system. Such that the original symmetry in nonmagnetic system, including the time-reversal symmetry are preserved in each spin channel. Given the states on path X-S is invariant under the combined Kramer-like operation $\mathcal{T}S_{2x}$, one also can derive that $(\mathcal{T}S_{2x})^2 = e^{-ik_x} = e^{-i\pi} = -1$ on this path. Hence, any band on this path is at least double degenerate due to this Kramer-like operation. Moreover, the path Y-S is the common invariant subspace for $\mathcal{T}S_{2y}$, with $(\mathcal{T}S_{2y})^2 = e^{-ik_y} = e^{-i\pi} = -1$ on this path, also giving rise to a double degeneracy at any k point on this path. Therefore, each band on these two paths is at least double degenerate, leading to nodal lines along these two boundaries of Brillouin zone.

Moreover, the two doubly degenerate bands overlapped with each other and formed a fourfold degenerate point ($D_1$) at the S point. Actually, such a fourfold degenerate $D_1$ point is a 2D Dirac point, which arises solely from the spin-down state (Figures 2b and 2d), making 2D $Mg_4N_4$ the first proposed genuine Dirac half-metal with 100% spin-polarization. For clarity, the three-dimensional plot of the spin-down Dirac point with fourfold degeneracy is shown in Figure 3a. We also employ the symmetry analysis to further understand the nature of the fourfold degenerate Dirac point at the S high-symmetry point as follows. We have mentioned that any point on path Y-S, including S point, is of double degeneracy due to the Kramer-like operation $\mathcal{T}S_{2y}$. We also note that S point is also a invariant point under inversion symmetry $\mathcal{P}$. Besides, one has $\mathcal{P}S_{2y} = e^{-ik_y}S_{2y}\mathcal{P}$, implying that these two operations anticommute each other at S ($\pi, \pi$) point, with $\{\mathcal{P}, S_{2y}\} = 0$. Therefore, states at S point are fourfold degenerate.

Then, we come to further understand the physics of the Dirac point in the spin-down channel of 2D $Mg_4N_4$. If the spin-orbit coupling is not considered, the spin-up and spin-down bands have no symmetry connecting them, making it possible to view the bands for each spin direction as a spinless system. Hence, we can choose a simple spinless four-band lattice model to show the key characteristics of the 2D Dirac point in LG 45.

A unit cell with one site (0,0,0) was considered in a spinless lattice model with LG 45,

and $p_x$ orbital was placed on this site. The spinless four-band lattice model can be written as:

$$\begin{pmatrix} H_{11} & H_{12} & 0 & H_{14} \\ & H_{11} & H_{14} & 0 \\ & & H_{11} & H_{12} \\ \dagger & & & H_{11} \end{pmatrix}, \quad (2)$$

where $H_{11} = 2r_2\cos(k_z)$, $H_{12} = 2t_1\cos\left(\frac{k_z}{2}\right)$, and $H_{14} = 2r_1\cos\left(\frac{k_x}{2}\right)$. Figure S4 depicts the band structure of the spinless lattice model (see Eq. (2)) along the Γ-X-S-Y-Γ paths. We set $e_1 = 0$, $r_1 = -0.115$, $r_2 = 0.005$, and $t_1 = -0.125$ for the bands, and a fourfold Dirac point appears at the S high-symmetry point. This simple model may serve as a starting point for investigating the 2D Dirac point in LG 45 in the future.

The Curie temperature, which dictates the thermal stability of the ferromagnetic ordering and is critical for spintronic device applications, determines the thermal stability of the ferromagnetic order. Then, we examine the Curie temperature of 2D Mg$_4$N$_4$ via Monto Carlo simulations based on the Heisenberg model with the single-ion anisotropy incorporated [46]. More details of the simulations can be found in the SM. The magnetic moment per unit cell and the magnetic susceptibility are shown in Figure 3b, from which the T$_C$ is found to be as high as 431 K in the 2D Mg$_4$N$_4$ monolayer.

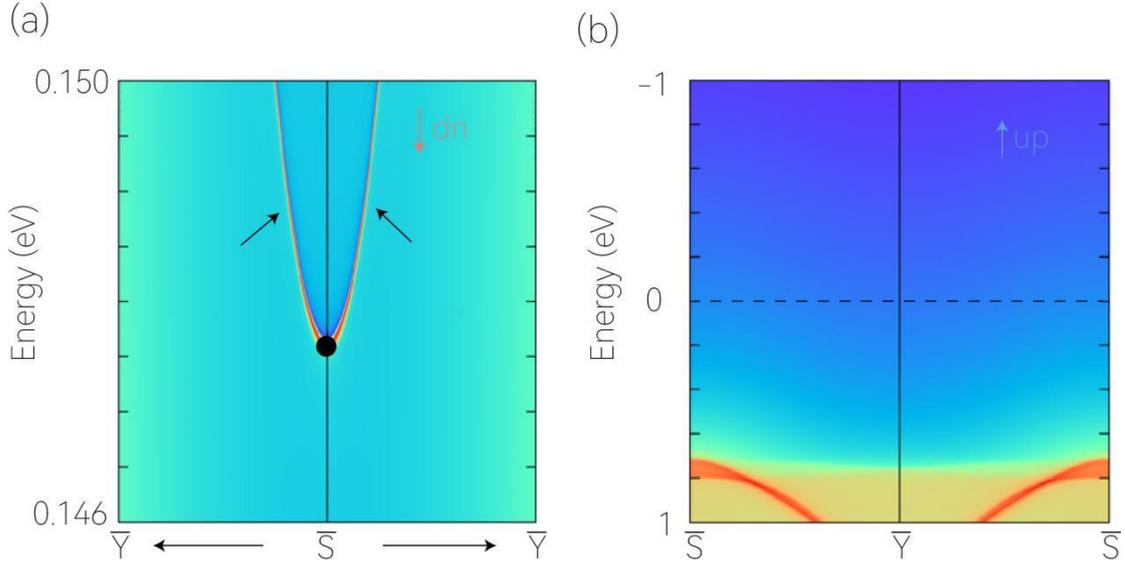

Figure 4. [010] projective spectra in the spin-down (a) and spin-up (b) channels. From (a), the fully spin-polarized edge states, connecting to the projection of the Driac point D$_1$, are apparent.

Furthermore, we studied the edge states of the Dirac point around the Fermi level along [010] (see Figure 1b). The results are shown in Figure 4. The location of the projected Dirac point in the spin-down channel is indicated by a black ball. The edge states are clearly connected by the projected Dirac point. Note that the edge states only appear in the spin-down channel, reflecting its fully spin-polarized nature. Following-up experiments using surface-sensitive sensors, such as electron energy loss spectroscopy and helium scattering, will benefit from such clean spin-polarized edge states.

Before closing this Letter, we would like to discuss the followings: (1) We further investigated the robustness of Dirac point in 2D $Mg_4N_4$ monolayer to the effect of spin-orbit coupling and uniaxial and biaxial strains (see the details in Figures S5-S7 of the SM). Results show that the Dirac point is robust to the spin-orbit coupling and uniaxial and biaxial strains for 2D $Mg_4N_4$ monolayer. (2) Besides the 2D $Mg_4N_4$ with LG No. 45, we also take 2D $Na_4O_4$ with LG 33 as an example to show it is a candidate for Dirac half-metal with a fourfold degenerate Dirac point at the S high-symmetry point, intrinsic magnetism, high Curie temperature, 100% spin-polarization, and robustness to the uniaxial and biaxial strains (see the details in Figures S8-S15 of the SM []). Moreover, as shown in Figure S12, we consider a unit cell with one site (0,0,0) and place the $p_x$ orbital on the (0,0,0) site. Thenm we build a spinless four-band lattice model to show the nature of the 2D Dirac point in LG 33 using Eq (2), where $H_{11} = e_1 + 2r_2\cos(k_z)$, $H_{12} = 2r_1\cos\left(\frac{k_z}{2}\right)$, and $H_{14} = 2t_1\cos\left(\frac{k_x}{2}\right)$.

In summary, we propose 2D $d^0$ ferromagnet $Mg_4N_4$ (with LG 45), a dynamically and thermodynamically stable material, is the first mentioned genuine Dirac half-metal with a fourfold degenerate Dirac point at the S high-symmetry point, intrinsic magnetism, high Curie temperature, 100% spin-polarization, and the robustness to the SOC and uniaxial and biaxial strains. Similar physics can also be found in 2D $d^0$ ferromagnet $Na_4O_4$ (with LG 33). As the first discovery of $d^0$ genuine Dirac half-metal in 2D, this Letter also offers simple models to show the key characteristics of the 2D Dirac points in LGs 45 and 33, which can be viewed as a starting point for predicting 2D genuine Dirac half-metals in both LGs.

1. Gibertini, M., Koperski, M., Morpurgo, A. F., & Novoselov, K. S. (2019). Magnetic 2D materials and heterostructures. *Nature nanotechnology*, *14*(5), 408-419.
2. Li, H., Ruan, S., & Zeng, Y. J. (2019). Intrinsic van der Waals magnetic materials from bulk to the 2D limit: new frontiers of spintronics. *Advanced Materials*, *31*(27), 1900065.
3. Hossain, M., Qin, B., Li, B., & Duan, X. (2022). Synthesis, characterization, properties and applications of two-dimensional magnetic materials. *Nano Today*, *42*, 101338.
4. Torelli, D., Moustafa, H., Jacobsen, K. W., & Olsen, T. (2020). High-throughput computational screening for two-dimensional magnetic materials based on experimental databases of three-dimensional compounds. *npj Computational Materials*, *6*(1), 158.
5. Gong, C., & Zhang, X. (2019). Two-dimensional magnetic crystals and emergent heterostructure devices. *Science*, *363*(6428), eaav4450.
6. H. Kumar, N. C. Frey, L. Dong, B. Anasori, Y. Gogotsi and V. B. Shenoy, ACS Nano, 2017, 11, 7648—7655
7. R. Nair, I.-L. Tsai, M. Sepioni, O. Lehtinen, J. Keinonen, A. Krasheninnikov, A. C. Neto, M. Katsnelson, A. Geim and I. Grigorieva, Nat. Commun., 2013, 4, 1—6
8. Z. Zhang, X. Zou, V. H. Crespi and B. I. Yakobson, ACS Nano, 2013, 7, 10475—10481
9. B. Huang, G. Clark, E. Navarro-Moratalla, D. R. Klein, R. Cheng, K. L. Seyler, D. Zhong, E. Schmidgall, M. A. McGuire, D. H. Cobden, W. Yao, D. Xiao, P. Jarillo-Herrero and X. Xu, Nature, 2017, 546, 270—273.
10. C. Gong, L. Li, Z. L. Li, H. W. Ji, A. Stern, Y. Xia, T. Cao, W. Bao, C. Z. Wang, Y. A. Wang, Z. Q. Qiu, R. J. Cava, S. G. Louie, J. Xia and X. Zhang, Nature, 2017, 546, 265—269.
11. Z. Fei, B. Huang, P. Malinowski, W. Wang, T. Song, J. Sanchez, W. Yao, D. Xiao, X. Zhu and A. F. May, Nat. Mater., 2018, 17, 778—782
12. Zhang, S. J., Zhang, C. W., Zhang, S. F., Ji, W. X., Li, P., Wang, P. J., ... & Yan, S. S. (2017). Intrinsic Dirac half-metal and quantum anomalous Hall phase in a hexagonal metal-oxide lattice. *Physical Review B*, *96*(20), 205433.
13. Chen, C. Q., Ni, X. S., Yao, D. X., & Hou, Y. (2022). Chern insulators and high Curie temperature Dirac half-metal in two-dimensional metal–organic frameworks. *Applied Physics Letters*, *121*(14), 142401.
14. Liu, Z., Liu, J., & Zhao, J. (2017). YN 2 monolayer: Novel p-state Dirac half metal for high-speed spintronics. *Nano Research*, *10*, 1972-1979.
15. Zhang, B., Sun, J., Leng, J., Zhang, C., & Wang, J. (2020). Tunable two dimensional ferromagnetic topological half-metal CrO2 by electronic correction and spin direction. *Applied Physics Letters*, *117*(22), 222407.
16. Ma, Y., Dai, Y., Li, X., Sun, Q., & Huang, B. (2014). Prediction of two-dimensional materials with half-metallic Dirac cones: Ni2C18H12 and

Co2C18H12. *Carbon*, *73*, 382-388.
17. Feng, Y., Wu, X., & Gao, G. (2020). High tunnel magnetoresistance based on 2D Dirac spin gapless semiconductor VCl3. *Applied Physics Letters*, *116*(2), 022402.
18. Yu, Y., Xie, X., Liu, X., Li, J., Peeters, F. M., & Li, L. (2022). Two-dimensional semimetal states in transition metal trichlorides: A first-principles study. *Applied Physics Letters*, *121*(11), 112405.
19. Hao, L., & Ting, C. S. (2016). Topological phase transitions and a two-dimensional Weyl superconductor in a half-metal/superconductor heterostructure. *Physical Review B*, *94*(13), 134513.
20. Yue, Z., Li, Z., Sang, L., & Wang, X. (2020). Spin-gapless semiconductors. *Small*, *16*(31), 1905155.
21. Wang, X. L. (2017). Dirac spin-gapless semiconductors: promising platforms for massless and dissipationless spintronics and new (quantum) anomalous spin Hall effects. *National Science Review*, *4*(2), 252-257.
22. Wang, X., Li, T., Cheng, Z., Wang, X. L., & Chen, H. (2018). Recent advances in Dirac spin-gapless semiconductors. *Applied Physics Reviews*, *5*(4), 041103.
23. Sun, Q., Ma, Y., & Kioussis, N. (2020). Two-dimensional Dirac spin-gapless semiconductors with tunable perpendicular magnetic anisotropy and a robust quantum anomalous Hall effect. *Materials Horizons*, *7*(8), 2071-2077.
24. Li, L., Kong, X., Chen, X., Li, J., Sanyal, B., & Peeters, F. M. (2020). Monolayer 1T-LaN2: Dirac spin-gapless semiconductor of p-state and Chern insulator with a high Chern number. *Applied Physics Letters*, *117*(14), 143101.
25. Wang, X., Cheng, Z., Zhang, G., Yuan, H., Chen, H., & Wang, X. L. (2020). Spin-gapless semiconductors for future spintronics and electronics. *Physics Reports*, *888*, 1-57.
26. Xing, J., Jiang, X., Liu, Z., Qi, Y., & Zhao, J. (2022). Robust Dirac spin gapless semiconductors in a two-dimensional oxalate based organic honeycomb-kagome lattice. *Nanoscale*, *14*(5), 2023-2029.
27. Nadeem, M., Hamilton, A. R., Fuhrer, M. S., & Wang, X. (2020). Quantum Anomalous Hall Effect in Magnetic Doped Topological Insulators and Ferromagnetic Spin-Gapless Semiconductors—A Perspective Review. *Small*, *16*(42), 1904322.
28. Yu, Y., Chen, X., Liu, X., Li, J., Sanyal, B., Kong, X., ... & Li, L. (2022). Ferromagnetism with in-plane magnetization, Dirac spin-gapless semiconducting properties, and tunable topological states in two-dimensional rare-earth metal dinitrides. *Physical Review B*, *105*(2), 024407.
29. Wang, Y., Jiang, J., Zou, J. J., & Mi, W. (2023). Spin-Gapless Semiconductors and Quantum Anomalous Hall Effects of Tetraazanaphthotetraphene-Based Two-Dimensional Transition-Metal Organic Frameworks on Spintronics and Electrocatalysts for CO2 Reduction. *ACS Applied Electronic Materials*, *5*(2), 1243-1251.
30. Yang, J., Zhou, Y., Dedkov, Y., & Voloshina, E. (2020). Dirac Fermions in Half-Metallic Ferromagnetic Mixed Cr 1− x MxPSe3 Monolayers. *Advanced Theory and Simulations*, *3*(12), 2000228.

31. Deng, Y., Yu, Y., Shi, M. Z., Guo, Z., Xu, Z., Wang, J., ... & Zhang, Y. (2020). Quantum anomalous Hall effect in intrinsic magnetic topological insulator MnBi2Te4. *Science*, *367*(6480), 895-900.
32. Fu, B., Ma, D. S., He, C., Zhao, Y. H., Yu, Z. M., & Yao, Y. (2022). Two-dimensional Dirac semiconductor and its material realization. *Physical Review B*, *105*(3), 035126.
33. Gong, J., Wang, J., Yuan, H., Zhang, Z., Wang, W., & Wang, X. (2022). Dirac phonons in two-dimensional materials. *Physical Review B*, *106*(21), 214317.
34. Zhang, Z., Wu, W., Liu, G. B., Yu, Z. M., Yang, S. A., & Yao, Y. (2023). Encyclopedia of emergent particles in 528 magnetic layer groups and 394 magnetic rod groups. *Physical Review B*, *107*(7), 075405.
35. Yang, S. A. (2016, June). Dirac and Weyl materials: fundamental aspects and some spintronics applications. In *Spin* (Vol. 6, No. 02, p. 1640003). World Scientific Publishing Company.
36. W. X. Ji, B. M. Zhang, S. F. Zhang, C. W. Zhang, M. Ding, P.-J. Wang, and R. Zhang, Nanoscale 10, 13645 (2018).
37. W. X. Ji, B. M. Zhang, S. F. Zhang, C. W. Zhang, M. Ding, P. Li, and P.-J. Wang, J. Mater. Chem. C 5, 8504 (2017).
38. Jin, L., Zhang, X., Liu, Y., Dai, X., Shen, X., Wang, L., & Liu, G. (2020). Two-dimensional Weyl nodal-line semimetal in a d 0 ferromagnetic K 2 N monolayer with a high curie temperature. *Physical Review B*, *102*(12), 125118.
39. Giannozzi, P., & Baroni, S. (2005). Density-functional perturbation theory. *Handbook of Materials Modeling: Methods*, 195-214.
40. Togo, A. & Tanaka, I. First principles phonon calculations in materials science. Scr. Mater. 108, 1–5 (2015).
41. Marx, D., & Hutter, J. (2000). Ab initio molecular dynamics: Theory and implementation. *Modern methods and algorithms of quantum chemistry*, *1*(301-449), 141.
42. J. P. Perdew, K. Burke, and M. Ernzerhof, Phys. Rev. Lett. 77, 3865 (1996).
43. J. Heyd, G. E. Scuseria and M. Ernzerhof, J. Chem. Phys., 2003, 118, 8207–8215.
44. Wang, J., Yuan, H., Liu, Y., Wang, X., & Zhang, G. (2022). Multiple dimensions of spin-gapless semiconducting states in tetragonal Sr 2 CuF 6. *Physical Review B*, *106*(6), L060407.
45. Ding, G., Wang, J., Yu, Z. M., Zhang, Z., Wang, W., & Wang, X. (2023). Single pair of type-III Weyl points half-metals: BaNiIO 6 as an example. *Physical Review Materials*, *7*(1), 014202.
46. B. Wang, X. Zhang, Y. Zhang, S. Yuan, Y. Guo, S. Dong and J. Wang, Mater. Horiz., 2020, 7, 1623–1630.